\documentclass[manuscript]{acmart}
\usepackage[nolist,nohyperlinks]{acronym}
\usepackage{amsmath}
\usepackage{lipsum} 
\usepackage{array}
\usepackage{varwidth}

\usepackage[skip=0.333\baselineskip]{caption}
\usepackage{subcaption}

\AtBeginDocument{%
  \providecommand\BibTeX{{%
    \normalfont B\kern-0.5em{\scshape i\kern-0.25em b}\kern-0.8em\TeX}}}

\copyrightyear{2023}
\acmYear{2023}
\setcopyright{rightsretained}
\acmConference[RecSys '23]{Seventeenth ACM Conference on Recommender Systems}{September 18--22, 2023}{Singapore}
\acmBooktitle{Seventeenth ACM Conference on Recommender Systems (RecSys '23), September 18--22, 2023, Singapore}
\acmDOI{TBD.}
\acmISBN{XXX-X-XXXX-XXXX-X/XX/XX}

\usepackage{multirow}

\begin{document}

\title{Track Mix Generation on Music Streaming Services using~Transformers}

\author{Walid Bendada}
\author{Théo Bontempelli}
\author{Mathieu Morlon}
\author{Benjamin Chapus}
\author{Thibault Cador}
\author{Thomas Bouabça}
\author{Guillaume Salha-Galvan}
\affiliation{
  \institution{Deezer}
  \city{}
  \country{France}
}
\renewcommand{\shortauthors}{W. Bendada, T. Bontempelli, M. Morlon, B. Chapus, T. Cador, T. Bouabça, G. Salha-Galvan}

\begin{abstract}
This paper introduces Track Mix, a personalized playlist generation system released in 2022 on the music streaming service Deezer. 
Track Mix automatically generates ``mix'' playlists inspired by initial music tracks, allowing users to discover music similar to their favorite content.
To generate these mixes, we consider a Transformer model trained on millions of track sequences from user playlists. In light of the growing popularity of Transformers in recent years, we analyze the advantages, drawbacks, and technical challenges of using such a model for mix generation on the service, compared to a more traditional collaborative filtering approach.
Since its release, Track Mix has been generating playlists for millions of users daily, enhancing their music discovery~experience~on Deezer.
\end{abstract}


\maketitle

\section{Introduction}
\label{s1}
The French music streaming service Deezer~\cite{deezerwebsite} provides unlimited access to a large catalog of 90 million music~tracks.
To help its 16~million active users navigate through this catalog and discover new content they might like, the service integrates a variety of large-scale music recommender systems~\cite{bendada2020carousel,briand2021semi,bontempelli2022flow,bendada2023scalable}. In this paper, we present the latest addition to these systems: Track Mix, a personalized playlist generation tool designed to enhance the music discovery experience~on~Deezer. 
Released in 2022 on the Deezer homepage, Track Mix generates ``mix'' playlists inspired by a selected initial music track.
This allows users to discover new tracks similar to their~favorite~ones~on~the~service.

To generate a playlist from an initial~track, we consider in this paper a Transformer~\cite{Vaswani2017AttentionNeed} trained on millions of track sequences obtained from user playlists.
We show that, while deploying this model in production entails facing important technical challenges, it improves most of our performance indicators in online A/B tests compared to a more traditional latent model for collaborative filtering~\cite{koren2015advances,bokde2015matrix}. Transformer-based mixes consistently result in longer listening times for all users. They are also associated with a significant increase in the number of ``collect'' actions (i.e., additions to the list of favorite tracks or to personal playlists) among new users. This is a valuable result, as facilitating the acquisition of preference information on new users contributes to addressing cold~start problems~\cite{briand2021semi,schedl2018current}. Nonetheless, this Transformer also reduces the number of collect actions for more regular users. We interpret this uneven performance in terms of popularity bias and user expectation regarding music discovery. 
This paper is organized as follows. Section~\ref{s21} introduces the Track Mix feature on Deezer. Section~\ref{s22} details the motivations, development, and deployment of our Transformer for mix generation. We analyze our online experiments on Deezer in Section~\ref{s23},~and~conclude~in~Section~\ref{s3}.

\section{Track Mix Generation on Deezer using Transformers}
\label{s2}

 \begin{figure*}[t]
  \centering
  \includegraphics[width=1.0\textwidth]{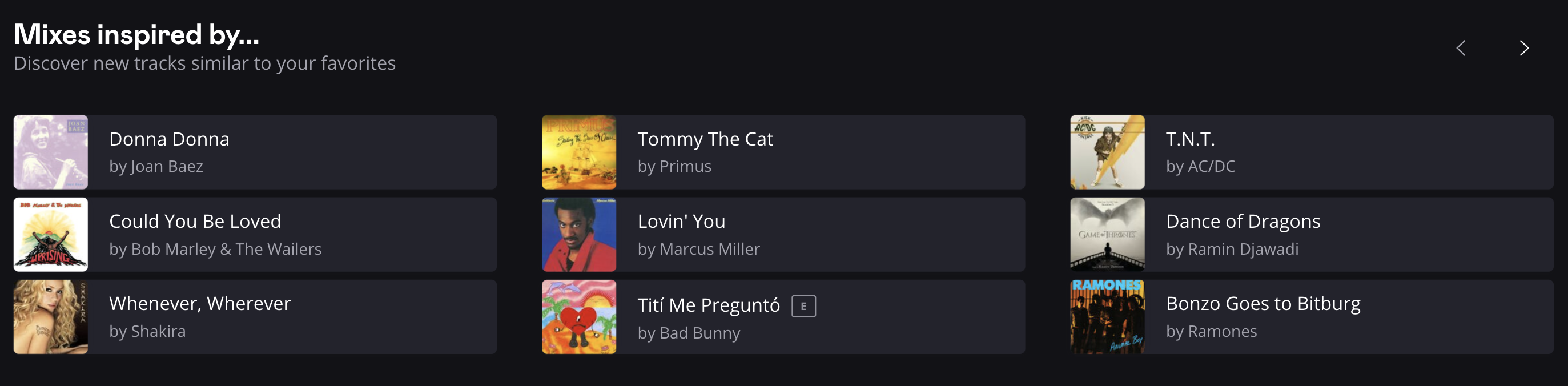}
  \caption{Interface of Track Mix on the website version of Deezer. The mobile app proposes a comparable feature. Track Mix presents a personalized and dynamic shortlist of music tracks among the ones previously liked and listened to by each user. A click on a track generates a mix playlist ``inspired by'' this track, with the aim of supporting users in discovering new tracks similar to~their~favorite~ones.}
  \label{fig:trackmix}
\end{figure*}

\subsection{Track Mix, a Personalized Playlist Generation System on Deezer}
\label{s21}

\subsubsection{The Track Mix Feature}
\label{s211}

\begin{figure*}[t]
  \centering
  \includegraphics[width=1.0\textwidth]{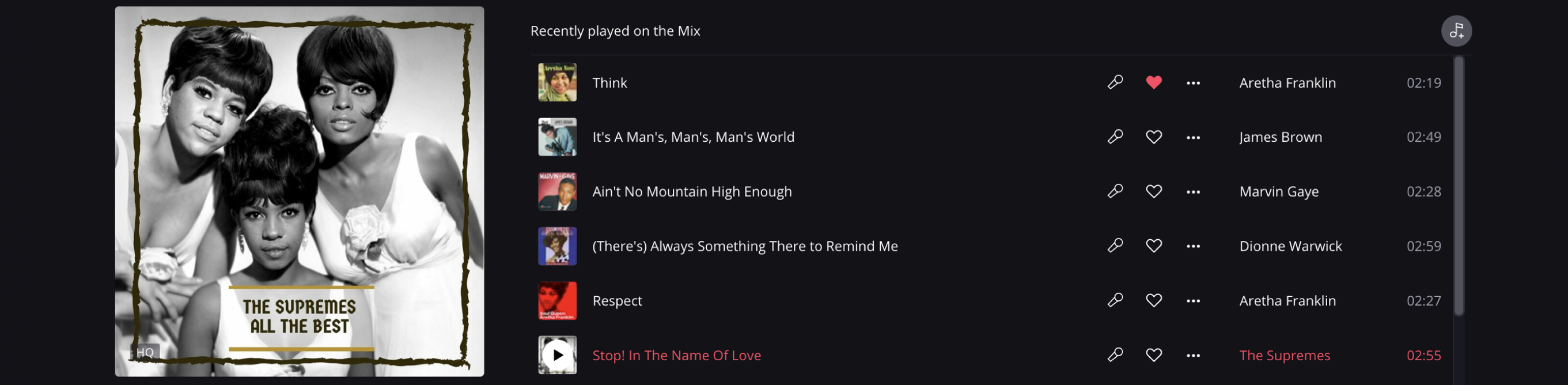}
  \caption{An example of the beginning of a Track Mix playlist on Deezer, generated with Mix-Transformer from ``Think'' by~Aretha~Franklin.}
  \label{fig:trackmix2}
\end{figure*}

Track Mix is a playlist generation system, accessible by millions of Deezer users since its worldwide release on the homepage of this service in 2022. 
As illustrated in Figure~\ref{fig:trackmix}, it materializes as a personalized shortlist of up to 12~music~tracks, selected\footnote{\label{footnote1} Technical details on the selection process for initial track and on playlist reordering are voluntarily omitted in this paper for confidentiality reasons.} from the ones previously liked or regularly listened to by each user. They are dynamically updated at each connection to the service.
As illustrated in Figure~\ref{fig:trackmix2}, a click on one of them generates a ``mix'' playlist composed of the selection and 39 other similar tracks. Besides serving as an online jukebox, Track~Mix aims to support users in discovering music similar to their favorite content.
Unlike the Flow algorithm~on~Deezer~\cite{bontempelli2022flow}, a~personalized radio mixing the user’s favorite tracks along with new recommendations, Track Mix does not automatically enforce the addition of favorites within playlists. Hence, Track Mix puts a stronger emphasis~on~music~discovery.

\subsubsection{A Baseline Model for Track Mix Generation}
\label{s212} 
To generate these playlists, Track Mix has been historically relying on ``Mix-SVD'', an internal latent model for collaborative filtering~\cite{koren2015advances,bokde2015matrix} that will act as a baseline for the Transformer from~Section~\ref{s22}.
By factorizing a pointwise mutual information matrix based on track co-occurrences in user playlists and lists of favorites, using singular value decomposition~(SVD)~\cite{banerjee2014linear}, this model learns vector representations of tracks in an embedding space where proximity reflects user~preferences~\cite{briand2021semi}. When a user selects an initial track in Track~Mix, this model identifies its closest neighbors in the embedding space, and reorders them using internal rules\textsuperscript{\ref{footnote1}} to generate playlists. Mix-SVD is suitable for large-scale production use on a service like Deezer. Embedding vectors of millions of tracks undergo weekly updates and are exported in a Cassandra cluster.  Computation services run on a Kubernetes cluster. We compute playlist generation operations on a Scala server. In particular, we use approximate nearest neighbors techniques~\cite{li2019approximate} for efficient similarity search, via a Golang application incorporating the Faiss~library~\cite{johnson2019billion}.

\subsection{Leveraging Transformers for Track Mix Generation}
\label{s22}

\subsubsection{Motivation}
\label{s221}
In this paper, we consider replacing Mix-SVD with a Transformer~\cite{Vaswani2017AttentionNeed}.
In its general formulation, the term Transformer refers to a family of neural architectures leveraging attention mechanisms~\cite{brauwers2021general} to process~sequential data.
In recent years, Transformers have emerged as a competitive approach for sequence modeling and generation in various domains~\cite{khan2022transformers,devlin-etal-2019-bert,brauwers2021general}. Besides the sequential nature of music playlists, we explore the use of these models for Track~Mix generation for two main reasons.
Firstly, they have~achieved promising results on recommendation tasks in recent research (see, e.g., SASRec and BERT4Rec~\cite{Kang2018Self-AttentiveRecommendation,Sun2019BERT4Rec:Transformer}). Secondly, a Transformer has already been successfully deployed on Deezer to improve automatic~playlist~continuation~(APC)~\cite{Chen2018RecSysContinuation,Zamani2019AnContinuation}, i.e., to better recommend lists of tracks for users to extend their own playlists (we refer to Bendada et al.~\cite{bendada2023scalable}~for~details). Our study in this paper aims to build upon these~successes. Nonetheless, previous work has also emphasized that the performance of Transformers on APC tends to decrease when the number of tracks in the playlist to extend diminishes~\cite{bendada2023scalable}. In the extreme case of the Track Mix feature, a Transformer would only have access to \textit{a single} initial track to generate a playlist. For this reason,  the empirical superiority of a Transformer on Track Mix generation compared to Mix-SVD still~needs~to~be~fully~demonstrated.

\subsubsection{Mix-Transformer}

\label{s222} The specific model we consider in our study, denoted ``Mix-Transformer'', is a Decoder-only Transformer \cite{Vaswani2017AttentionNeed} with one hidden layer.
Our decision to solely retain the Decoder component is driven by the popularity of Generative Pre-trained Transformers (GPT), a family of Decoder-only Transformers with state-of-the-art performances on sequence generation~\cite{brown2020language,radford2018improving,casola2022pre}. 
We train Mix-Transformer on a playlist~completion~task, using millions of track sequences obtained from user playlists on Deezer. Formally, we denote by~$\mathcal{T}$ the set of tracks from the Deezer catalog and by $z_{t} \in \mathbb{R}^{d}$ some embedding vector representing each track $t \in \mathcal{T}$, with $d \in \mathbb{N}^*$. A playlist of length $l \in \mathbb{N}^+$ is a sequence of  $l$ distinct tracks from~$\mathcal{T}$. For each playlist $p$ of length $l > 1$ and $k \in \{1, \dots, l-1\}$, we create a sub-playlist $p_{:k}$ consisting of the $k$ first elements of~$p$, and a sub-playlist $p_{(l-k):}$ consisting of its $l-k$ last elements. We train Mix-Transformer to associate, to each $p_{:k}$, an embedding vector $z_{p_{:k}} \in \mathbb{R}^d$ whose inner product nearest neighbors should be the embedding vectors of tracks in $p_{(l-k):}$. For this purpose, we minimize the same logistic loss as our previously mentioned APC model~\cite{bendada2023scalable}, using tracks randomly picked from~$\mathcal{T} \setminus p$ as negative samples, and initial track embedding vectors retrieved from Mix-SVD. 
Finally, to generate a mix playlist in Track Mix, the trained Mix-Transformer treats the selected initial track as a 1-track playlist, computes its embedding vector, and then identifies and reorders\textsuperscript{\ref{footnote1}}~its~closest~neighbors.

%


\subsubsection{Deployment}
\label{s223} 

We stress that deploying Mix-Transformer on Deezer entails overcoming~engineering~challenges. 
Our internal tests have revealed that directly replacing Mix-SVD with Mix-Transformer in Track Mix would result in four times longer inference times for mix generation.
This latency would be deemed unacceptable in production, as it would be noticeable to users. Additionally, the maximum throughput, i.e., the number of users that could be served simultaneously, would severely deteriorate. To obtain latency and throughput levels comparable to Mix-SVD, without incurring additional infrastructure costs (e.g., without adding~GPUs), we closely follow the recently proposed ``represent-then-aggregate'' framework~\cite{bendada2023scalable} for scalable APC.
We leverage ONNX model merging~\cite{onnx} to integrate all operations from data processing to track ranking into the deployed Mix-Transformer. This reduces its infrastructural complexity and speeds up inferences.
Lastly, and perhaps most importantly, we dynamically quantize~\cite{onnx-dynamic} ONNX models. Our tests have demonstrated that this quantization significantly reduces computation costs, with a~minor~impact~on~performances. 

\subsection{Online Evaluation on Deezer}
\label{s23}

\begin{table*}[t]\centering
\caption{Online A/B test on Track Mix: relative performance of Mix-Transformer compared to Mix-SVD. Results are computed~on French Deezer Premium users in March-April 2023. All improvements/declines are statistically significant at the 1\% level (p-value~<~0.01). We note that, although this table focuses on France, i.e., Deezer's home market, we obtain comparable results in other countries.}
\label{table:results}
\resizebox{1.0\textwidth}{!}{
\begin{tabular}{cc|r||c||c|c}
\toprule
&  & & \multicolumn{3}{c}{\textbf{Relative performance of Mix-Transformer}} \\
\multicolumn{2}{c|}{\textbf{Metric Type}} & \multicolumn{1}{c||}{\textbf{Metric Description}}& \small{Results on} & \small{Results on users with} &  \small{Results on users with} \\
& & & \small{all users} & \small{seniority $\leq$ 30 days} &  \small{seniority $>$ 30 days} \\
\midrule
\midrule 
\multirow{3}{*}{User-centric}& \multirow{2}{*}{Listening time} & Median daily listening time per user & + 6.84\% & + 7.75\% & + 6.83\% \\
 &  &Average daily number of streams per user & + 4.23\% & + 5.07\% & + 4.26\% \\
  \cline{2-2}
& Music discovery & Average daily number of collect actions\textsuperscript{\ref{footnote2}} per user & - 2.48\% & + 20.26\% & - 3.79\% \\
\midrule
 \midrule
\multirow{3}{*}{Playlist-centric}& \multirow{2}{*}{Listening time}  & Average listening time per generated playlist & + 5.12\% & + 6.62\% & + 4.98\% \\
 &  &  Percentage of generated playlists listened to for $>$15min  & + 6.56\% & + 8.69\% & + 6.20\% \\
 \cline{2-2}
& Music discovery & Average number of collect actions\textsuperscript{\ref{footnote2}} per generated playlist & - 1.98\% & + 24.10\% & - 3.88\% \\
\bottomrule
\end{tabular}
\label{sequenceevaluation}
}
\end{table*}

\subsubsection{Setting}
\label{s231}
We now present the online A/B test we conducted on millions of Deezer users in March and~April~2023.
During this test, we used Mix-Transformer to generate Track Mix playlists for a randomly selected cohort of test users, and Mix-SVD~for~others.
Table~\ref{table:results} reports the relative performance of~Mix-Transformer~compared~to~Mix-SVD.
Four metrics compare listening times for each group, and two metrics evaluate the music~discovery aspect via collect actions\footnote{\label{footnote2} The number of ``collect'' actions is the number of recommended tracks that users added to their list of favorites or to their personal playlists.}.

\subsubsection{Results}
\label{s232} 

We observe that using Mix-Transformer enhances the listening times of Track~Mix~playlists, according to all four metrics under consideration. This positive result indicates a higher usage of the feature on Deezer. However, the number of collect actions simultaneously diminishes in the Mix-Transformer cohort.
A close examination of results reveals that this decrease primarily affects users with over a month of activity (seniority~$>$~30~days). Conversely, Mix-Transformer boosts collect actions for new~users (seniority~$\leq$~30~days).  
Hence, using Mix-Transformer would facilitate the acquisition of preference information on these new users, which could benefit all usage-based recommender systems on Deezer by helping to overcome cold start issues~\cite{briand2021semi}. We note that Mix-Transformer tends to recommend more popular tracks than Mix-SVD. This increased mainstreamness might explain the lower collection levels among regular users. Indeed, they might have already liked these popular tracks. Moreover, they might have different expectations regarding music discovery. Having an already established library of favorite content, they might be more open to and even looking for specialized/niche recommendations.
While Table~\ref{table:results} supports these assumptions, more investigations  will be required in future work for confirmation. Our test also encourages the development of an improved Mix-Transformer addressing popularity biases~\cite{schedl2018current} on regular users, e.g., by controlling popularity levels in the training dataset~and~the~optimized~loss.

\section{Conclusion}
\label{s3}

Since its release in 2022, Track Mix has been generating mix playlists for millions of Deezer users daily. At the time of writing,  Mix-SVD remains in use on the service for mix generation. However, our online A/B test prompts us to consider adopting Mix-Transformer for new users in the near future. Despite being more challenging to deploy, this method showcases a promising performance with these users. Our test also opens up interesting avenues for future research on Transformer-based mix generation, to further enhance the music discovery experience of~our~more~regular~users.



\section*{Speaker Bio}
\label{bio}

\textbf{Guillaume Salha-Galvan} is a research coordinator at Deezer, where he conducts fundamental and applied research projects on music recommendation.  He holds a Ph.D. in Computer Science from École Polytechnique~in~France.

\bibliographystyle{ACM-Reference-Format}
\bibliography{references}

\end{document}